\documentclass[prb,amsmath,amssymb,floatfix]{revtex4}
\usepackage{graphicx}
\usepackage{float}
\usepackage{bm}
\usepackage[utf8]{inputenc}
\usepackage{color}
\begin{document}
\title{Three-dimensional patchy lattice model for empty fluids}
\author{N. G. Almarza}
\affiliation{Instituto de Qu{\'\i}mica F{\'\i}sica Rocasolano, CSIC, Serrano 119, E-28006 Madrid, Spain }

\author{J. M. Tavares}
\affiliation{ Centro de F{\'\i}sica Te\'orica e Computacional, Universidade de Lisboa,
Avenida Professor Gama Pinto 2, P-1649-003 Lisbon, Portugal and
Instituto Superior de Engenharia de Lisboa, Rua Conselheiro Em\'{\i}dio Navarro 1, 
P-1950-062 Lisbon, Portugal}

\author{E. G. Noya}
\affiliation{ Instituto de Qu{\'\i}mica F{\'\i}sica Rocasolano, CSIC, Serrano 119, E-28006 Madrid, Spain }

\author{M. M. Telo da Gama}
\affiliation{Centro de F{\'\i}sica Te\'orica e Computacional, Universidade de Lisboa,
Avenida Professor Gama Pinto 2, P-1649-003 Lisbon, Portugal and
Departamento de F{\'\i}sica, Faculdade de Ci{\^e}ncias, Universidade de Lisboa,
Campo Grande, P-1749-016 Lisbon, Portugal}

\date{\today}
\begin{abstract}
The phase diagram of a simple model with two patches of type $A$ and ten patches
of type $B$ ($2A10B$) on the face centred cubic lattice 
has been calculated by simulations and theory. Assuming that there is
no interaction between the $B$ patches the behavior of the system 
can be described in terms of the ratio of 
the $AB$ and $AA$ interactions, $r$.
Our results show that, similarly to what
happens for related off-lattice and two-dimensional lattice models, the 
liquid-vapor phase equilibria exhibits reentrant behavior for some values
of the interaction parameters. 
However, for the model studied here the liquid-vapor phase
equilibria occurs for values of $r$ lower than $ \frac{1}{3}$, a
threshold value which was previously thought to be universal for $2AnB$ models.
In addition, the theory predicts that below $r=\frac{1}{3}$ (and above a new 
condensation threshold which is $< \frac{1}{3}$) 
the reentrant liquid-vapor equilibria is so extreme that it exhibits 
a closed loop with a lower critical point, a very unusual behavior in single-component systems.
An order-disorder transition is also observed at higher densities than the liquid-vapor equilibria, which
shows that the liquid-vapor reentrancy occurs in an equilibrium region of the phase diagram.
These findings may have implications in the understanding of the condensation 
of dipolar hard spheres given the analogy between that system and the $2AnB$ models considered here.

\end{abstract}
\maketitle

\section{Introduction}

Advances in the fabrication of nanometer-to-micrometer sized particles enable
tailoring their size, shape and interactions,
but their organization into complex structures remains a challenge. Self-
assembly is an appealing route of this bottom-up approach as the structure of
the clusters is tunable through the anisotropy of the particle shapes and
interactions. In addition, the strongly anisotropic interactions prevent the
clustering that drives condensation and have been shown to lead to novel
macroscopic behavior, including empty liquids, optimal networks and
equilibrium gels \cite{Glotzer_solomon,Pawar10,Bianchi11}.

Patchy particle models with dissimilar patches ($A$ and $B$) were introduced
in this context \cite{Tavares,Tavares_MP09} and allow a unique control of the effective
valence through the temperature $T$. In three-dimensional (3D) off-lattice
models consisting
of particles with two types of patches, $A$ and $B$, with no interaction between the $B$
patches, the topology of the liquid-vapor diagram is determined by the ratio
between the $AB$ and the $AA$ interactions, $r=\epsilon_{AB}/\epsilon_{AA}$. As
$r$ decreases in the range $\frac{1}{3}< r <
\frac{1}{2}$, the low-temperature liquid-vapor coexistence region also
decreases \cite{Russo2}. The binodal exhibits a reentrant shape with the
coexisting liquid density vanishing as the temperature approaches zero
\cite{Russo2,Russo1}. Below $r = \frac{1}{3}$ condensation is no
longer observed, and above $r = \frac{1}{2}$ there is no 
reentrant behavior \cite{Tavares2010}.

Both the scaling of the vanishing critical parameters and the reentrant phase
behavior are predicted correctly
by Wertheim's thermodynamic first-order perturbation 
theory \cite{Russo2,Russo1,Wertheim1984,Wertheim1984b,Wertheim1986,Wertheim1986b}.
The theory also reveals that the reentrant phase behavior is
driven by the balance of two entropic contributions: the higher entropy of the
junctions and the lower entropy of the chains in the (percolated) liquid phase,
as suggested a decade ago on the basis of a hierarchical theory of network
fluids \cite{Tulsty2000}.

In a previous paper we considered the $2A2B$ model consisting of particles
with four patches, two of type $A$ and two of type $B$, on the square lattice and investigated
its global phase behavior by simulations and theory \cite{AlmarzaJCP2011}. We have set
the $BB$ interaction to zero and calculated the phase diagram, as a function
of $r$. We found that, in the same range of parameters as in the
3D off-lattice models, the liquid-vapor diagram exhibits a reentrant shape \cite{AlmarzaJCP2011},
with a region where the system exhibits empty fluid behavior \cite{bianchi}.
In addition, below $r = \frac{1}{3}$ condensation ceases to
exist, while the reentrant regime disappears for $r > \frac{1}
{2}$, in line with the results for 3D off-lattice models and the predictions
of Wertheim's theory \cite{Tavares,Tavares_MP09,Russo2,Russo1}. When $r < \frac{1}{3}$ the
gain in entropy resulting from the $AB$ bonds does not balance the
loss in energy of the favored $AA$ bonds, in line with the simulation results
\cite{Russo2,Russo1,AlmarzaJCP2011}. This led us to suggest that the thresholds predicted by
Wertheim's theory are exact and universal, i.e. independent of the
dimensionality of the system and of the lattice structure\cite{AlmarzaJCP2011}.

Computer simulations of the phase diagram of these models become more and
more demanding as $r$ decreases. One of the reasons is the
rapid increase in the size of the voids of the empty coexisting liquid, as the
temperature decreases. Simulations of larger and larger systems are thus required to
obtain reliable results, rendering the computation of phase equilibria in the
neighboorhood of $r=\frac{1}{3}$ prohibitive \cite{AlmarzaJCP2011}.

In this paper we present results that clarify the role of  the $\frac{1}{3}$ threshold and 
the universality of the empty fluid regime reported earlier. We consider
a model consisting of particles with twelve bonding sites (``patches''), two
of type $A$ and ten of type $B$, on the face centered cubic lattice,
and investigate its global phase behavior by simulation and theory.
As before we set the interaction between the $B$ patches to zero. The
potential energy of this $2AnB$ class of models, is given by:
\begin{equation}
{\cal U} = - \epsilon_{AA} {\cal  N}_{AA} - \epsilon_{AB} {\cal N}_{AB} = 
- \left( {\cal N}_{AA} + r {\cal N}_{AB} \right) \epsilon_{AA};
\label{ugeneric}
\end{equation}
where $\epsilon_{AB}$ and $\epsilon_{AB}$ are positive, and ${\cal N}_{AA}$
and ${\cal N}_{AB}$ are the number of $AA$ and $AB$ bonds, respectively.

The model is a 3D counterpart of the $2A2B$ model on the square lattice
\cite{AlmarzaJCP2011}. We develop efficient sampling algorithms to simulate the
empty liquid at low temperatures and observed liquid-vapor coexistence in
systems with $r < \frac{1}{3}$. 

We establish a new non-universal threshold
$r_{m} < \frac{1}{3}$ for liquid-vapor equilibrium (LVE) and Wertheim's first-order perturbation theory 
suggests the existence of a new regime where the reentrant behavior is so extreme that the low 
temperature binodal closes at a lower critical point. Using an asymptotic expansion of the theory, we calculate the 
new threshold and show that the closed loop regime exhibits some degree of universality as it depends only on the 
number of $B$ patches. 
The conditions to observe the closed miscibility loop are met in lattice models with a large
number of $B$ patches, such as the $2A10B$ lattice model, but are unlikely to occur in continuum 
models with the same number of patches \cite{Russo2,Russo1}. The existence of a threshold $r_{m}< \frac{1}{3}$ 
for LVE and its degree of {\it universality} have physical relevance, as it has been used to address the liquid-vapor condensation of dipolar hard spheres \cite{TavaresMP2011}. 

In more recent work, the absence of liquid-vapor coexistence of dipolar fluids was related to ring formation \cite{RovigattiPRL2011}, a feature which is not described by Wertheim's first-order perturbation theory on which the reported thresholds are based. A related study of the generic phase diagram of $2A4B$ models on the triangular lattice with $r > \frac{1}{3} $, reports that orientational correlations between the $A$ patches that promote 
ring formation have a profound effect on the phase equilibria \cite{Almarza_closedloop}. No phase coexistence is observed 
when short rings are formed. Closed miscibility loops are found if larger rings are formed while the usual reentrant 
behavior is observed if no rings are formed. Somewhat surprisingly the same regimes are reported in 
this work based on Wertheim's first order perturbation theory, which does not account for ring 
formation, where  $r$
is the 
control parameter. 
The orientation of the $AA$ bonds that promotes ring formation, on the triangular lattice, has an effect on the topology 
of the phase diagram which is similar to the effect of decreasing the $AB$ interaction in a system without rings but with a large volume available for the formation of $AB$ bonds. Beyond 
this observation, the relation between the generic phase diagrams of $2AnB$ systems, with and without rings, is an
important open question that will be addressed in future work.

The remainder of this paper is arranged as follows: In Sec. \ref{sec:3DModel}
we describe the model. In Sec. \ref{sec:Simulation} we describe the Monte
Carlo techniques used to compute the phase diagrams. In Sec. \ref{sec:Results}
we present the simulation results while in Sec. \ref{sec:theory} we carry out the theoretical analysis. Finally, in Sec.
\ref{sec:Discussion} we discuss these and previous results and the
perspectives for future work.

\section{Three dimensional model}
\label{sec:3DModel}

We consider a face-centered-cubic (FCC) lattice. Sites on the lattice can be either empty or occupied
by, at most, one particle. The particles carry twelve patches: two of them of type $A$, and ten of
type $B$. The patches on each particle are oriented in the directions linking the site with its
twelve nearest neighbors (NN). The angle between the two $A$ patches is 180 degrees and the line 
between these patches defines the orientation of the particle.
On the FCC lattice, the particles are oriented along one of the $q=6$ equivalent directions $s_i \equiv s({\vec r}_i)=1,2,\cdots,q $ and thus a lattice site has $q+1$ possible states ($q=6$ orientations plus the additional empty state $s_i=0$).
The grand canonical Hamiltonian can be written as ${\cal H}= {\cal U}- \mu N$, with
$\mu$ being the chemical potential and $N$ the number of occupied sites:
\begin{equation}
  N =  \sum_{i=1} ^M \left[  1- \delta_{0,s_i}\right] ;
\end{equation}
where $\delta$ is Kronecker's delta function and $M$ the total
number of lattice sites. The potential energy ${\cal U}$ can be written as:
\begin{equation}
{\cal U} = \sum_{i=1}^M \sum_{k=1}^{q}  
V_2[s({\vec r}_i), s({\vec r}_i + {\vec \alpha}_k)]  
\label{u1}
\end{equation}
where $V_2$ is the interaction between pairs of NN sites, which
depends on the states of the sites and the direction of ${\vec \alpha}_k$ (the vector linking the two sites).
The pair interaction is defined as:

\begin{equation}
V_2[s({\vec r}_i),s({\vec r}_i+{\vec \alpha}_k)] = 
\left\{\begin{array}{llllll}
- \epsilon_{AA}& ; & \textrm {If:  } s({\vec r}_i) = k; & s({\vec r}_i+{\vec \alpha}_k) = k \\
- \epsilon_{AB}& ; & \textrm {If:  } 
s({\vec r}_i) \ne  k; & 
s({\vec r}_i) \ne  0;&
 s({\vec r}_i+{\vec \alpha}_k) = k \\
- \epsilon_{AB}& ; & \textrm {If:  } s({\vec r}_i) = k; &
 s({\vec r}_i+{\vec \alpha}_k) \ne k ; &
 s({\vec r}_i+{\vec \alpha}_k) \ne 0 ; \\
- \epsilon_{BB}& ; & \textrm {If:  } 
s({\vec r}_i) \ne  k; &
s({\vec r}_i) \ne  0; &
 s({\vec r}_i+{\vec \alpha}_k) \ne k ; &
 s({\vec r}_i+{\vec \alpha}_k) \ne 0 ; \\
- 0 & ; & \textrm {If:  } 
s({\vec r}_i) = 0 \\
- 0 & ; & \textrm {If:  } 
 s({\vec r}_i+{\vec \alpha}_k) = 0 \\
\end{array}
\right.
\label{u2}
\end{equation} 

In short, interactions are defined between two NN sites, $i$ and $j$,
if both are occupied, with a magnitude that depends on the types
of patches on each site pointing to the other one. We take $\epsilon \equiv \epsilon_{AA}$
as the energy scale and in line with previous work set:  $\epsilon_{BB}=0$.

The $2A10B$ model with $\epsilon_{BB}=0$ and $r = \epsilon_{AB}/\epsilon \le \frac{1}{2}$ may exhibit 
two phase transitions (as in 2D \cite{AlmarzaJCP2011}) which are
interpreted as a liquid-vapor transition, and an order disorder transition,
the latter occurring at a higher density when both transitions occur at a given temperature.

\section{Simulation Methods}
\label{sec:Simulation}

We have adapted the methodology used in 2D \cite{AlmarzaJCP2011} to the 3D lattice model. In addition 
we have built up cluster algorithms to enhance the sampling procedures. In 3D the order-disorder transition 
was found to be discontinuous, which simplifies the calculation of the phase diagram.
We used cubic cells of different sizes and periodic boundary conditions. The number of sites of a given 
system is $M=4 L^3$, with $L$ an integer.

\subsection{Liquid-vapor transition}

We have considered different values of $r$
in the range $[0.30,0.50]$.
The basic protocol to compute the liquid-vapor equilibria (LVE) was as follows:
First we used 
Wang-Landau multicanonical (WLMC)
procedures  \cite{Lomba,Ganzenmuller,HSDHS,almarza2009}, supplemented with a finite-size scaling 
analysis\cite{Lomba,Perez,HSDHS,almarza2009}  to compute 
the value of the
chemical potential of the transition at some subcritical temperature, $T_0$.  This
temperature is chosen not too close to the critical temperature of the LVE to
find a clear separation between the two phases, but not too far to avoid the sampling difficulties
that appear at low temperatures. 
For different system sizes, $L$, we computed the value of the chemical potential
that maximized the density fluctuations of the system, $\mu_e(L,T_0)$; this
is considered the finite-size estimate for the LVE at temperature $T_0$. Then,
the value in the thermodynamic limit, $\mu_e(T_0)$, is obtained by fitting the results
to:
\begin{equation}
\mu_e(L,T_0) = \mu_e(T_0) + a L^{-d};
\label{extrapolation}
\end{equation}
 where $d$ is the spatial dimensionality of the system, $d=3$. 
The range of $L$ required to get a reliable estimate of $\mu_e(T_0)$ depends
dramatically on the value of $r$. As one reduces $r$, larger
values of $L$ are required. The results were checked
by performing a fully independent estimate by means of thermodynamic integration
techniques using relatively large system sizes. Details of the implementation
of these techniques will be described later
in the paper.

Once we have estimated a reference point for the LVE: $[T_0,\mu_e(T_0)]$ 
a Gibbs-Duhem integration procedure was carried out to draw the binodal lines, using a methodology similar
to that described in previous papers \cite{AlmarzaJCP2011,Noya,Hoye}. The only relevant difference is that
in the grand canonical simulation of each phase, we incorporate collective
moves via the cluster algorithms described in the Appendix.

\subsubsection{Wang-Landau multicanonical methodology}

The basic strategy of the multicanonical procedures is to sample in a single run
the properties of the system with different numbers of particles ($N$). The procedure can
be related to a simulation in the grand canonical ensemble (GCE). In both cases the probability of 
a given configuration, ${\bf S}_{N}$ (with $N$ occupied sites), can be written as:
\begin{equation}
P({\bf S}_{N}) = \omega_0(N) \exp \left[ - \beta {\cal U} \left ({\bf S}_N \right) \right],
\end{equation}
where $\beta = (k_B T)^{-1}$, $T$ being the temperature, and $k_B$ the Boltzmann's constant.
In the GCE simulation the weighting factor is taken to be:
$\omega_0(N) \propto   \exp \left( \beta \mu N \right)$
whereas in the WLMC method, $\omega_0(N) $ is chosen to obtain a flat histogram of the density $P(N) \simeq
1/(1+N_{max}-N_{min}); \forall N \in [N_{\min},N_{max}]$, which in practice implies
$\omega_0 (N)\propto \exp \left[ \beta F(N,M,T) \right] $, $F$ being the Helmholtz thermodynamic
potential. In order to compute $\omega_{0} (N)$ in the WLMC procedure an equilibration scheme based on the
Wang-Landau strategy
\cite{WL_PRL2001,WL_PRE2001} is carried out, where the values of $\omega_{0}(N)$ can vary through the first part
of the simulation\cite{Lomba}. Once $\omega_{0}(N)$ is obtained, the equilibrium simulation (fixed $\omega_{0}(N)$) is run, 
and values of the different properties are computed as a function of $N$. These results can be used to determine
phase equilibria, and using appropriate reweighting techniques to locate the critical points.

The WLMC simulations include two types of moves: insertion and deletion of particles
on the lattice. In most cases ($r \ge 0.31$) we used a simple non-biased algorithm: particles to be removed, and the position and orientation of the inserted particles were chosen at random (among non-occupied lattice sites). 
Taking into account detailed balance\cite{AllenTildesleyBook},
the acceptance criteria for these moves can be written as:
\begin{equation}
A({\bf S}_{N+1}|{\bf S}_N) = \min \left\{ 1, \frac{\omega_{0}(N+1)}{\omega_{0}(N)}
e^{-\beta \Delta U_{\text{ins}} }
\frac
{  (M-N) q }
{N+1 }
\right\};
\label{acp}
\end{equation}
\begin{equation}
A \left( {\bf S}_{N-1}|{\bf S}_{N} \right) = 
\min \left\{ 1, 
\frac{\omega_{0}(N-1)}{\omega_{0}(N)}
e^{-\beta \Delta U_{\text{del}} }
\frac
{N}
{ (M-N+1) q}
\right\};
\label{acm}
\end{equation}
where $\Delta U_{\text{ins}}$ and $\Delta U_{\text{del}}$ are respectively the variations of the
energy when inserting and deleting one particle.
The factor $q$ in Eqs. (\ref{acp}-\ref{acm}) takes into account the
$q$  possible orientations of one particle, whereas the $M-N$ like terms
are the number of empty sites where the insertions can be attempted.

For the systems
 with the lowest values of $r$ ($r <  0.31$)
(and at low temperatures),
the previous scheme was found to loose efficiency. Then, in addition to the random insertions
and deletions described above, we include biased insertion/deletion moves. 
In the biased moves only insertions (deletions) that lead to the formation (destruction)
of, at least, one $AA$ bond are considered. The insertions are exclusively tried
in empty sites to which at least one non-bonded $A$ patch is pointing. We choose
at random with equal probabilities one of those non-bonded $A$ patches, and
perform the trial insertion of the new particle at the site to which the chosen patch
is pointing to, with the orientation that guarantees the formation of an $AA$ bond
with that patch. Deletions are carried out following the symmetric move, i.e.,
we choose at random, with equal probability, one of the $A$-patches that participates
in an $AA$ bond, and the particle to which the $A$ patch points to is removed.
Taking into account the conditions of detailed balance, the acceptance criteria of these moves 
can be written as:
\begin{equation}
A^{\text{(bias)}}({\bf S}_{N+1}|{\bf S}_N) = \min \left\{ 1, \frac{\omega_{0}(N+1)}{\omega_{0}(N)}
e^{-\beta \Delta U_{\text{ins}} }
\frac
{ {\cal N}_{A0} ({\bf S}_N)}
{2{\cal N}_{AA}({\bf S}_{N+1}) }
\right\};
\end{equation}
\begin{equation}
A^{\text{(bias)}} \left( {\bf S}_{N-1}|{\bf S}_{N} \right) = 
\min \left\{ 1, 
\frac{\omega_{0}(N-1)}{\omega_{0}(N)}
e^{-\beta \Delta U_{\text{del}} }
\frac
{2{\cal N}_{AA}({\bf S}_{N}) }
{ {\cal N}_{A0} ({\bf S}_{N-1})}
\right\};
\end{equation}
where ${\cal N}_{AA}$ is the number of $AA$ bonds in the system, and ${\cal N}_{A0}$ is
the number of $A$ patches that point to an empty site. Before attempting a MC step
we choose with equal probabilities whether to use fully random or biased moves.

\subsubsection{Critical points}

The critical points of the LVE were estimated using the Wang-Landau multicanonical
algorithm. After preliminary short calculations to locate approximately
the critical temperature we run, as above, simulations for different system sizes.
By storing histograms of the mean values of the potential energy: $<U>$
and its square $<U^2>$ as a function of the number of particles $N$ we can
apply a reweighting scheme to the results to estimate the thermodynamics
at temperatures close to that where the multicanonical simulation is carried out.\cite{Lomba}

We proceed to calculate the pseudo-critical points $(\mu_c^{(L)},T_c^{(L)})$, with $\mu_c^{(L)} \equiv \mu_e(T_c^{(L)},L)$. 
We compute the cumulants $g_4(L,T,\mu_e)$ defined by $g_4 \equiv  m_4/m_2^2$, where $m_k$ are the $k^{th}$ order moments of the density distribution function $P(\rho|T,\mu_e,L)$:
\begin{equation}
m_k = < \left( \rho - \bar{\rho} \right)^k > ;
\end{equation}
and $\bar{\rho}$ is the average of the density. The pseudocritical point satisfies 
\cite{Wilding,Perez,Lomba,AlmarzaJCP2011}:
\begin{equation}
g_4(L,T_c^{(L)}, \mu_c^{(L)}) = g_4^{(c)};
\end{equation}
where $g_4^{(c)}$ is a constant that depends on the universality class and on the boundary conditions. The liquid-vapor critical point of $2AnB$ models is expected to be in the 3D Ising universality class, with $g_4^{(c)} \simeq  1.604 $ \cite{Blote}. 
The pseudocritical densities are taken as:
\begin{equation}
\rho_c(L) = \bar{\rho}(L,T_c^{(L)},\mu_c^{(L)}).
\end{equation}
The estimates for the critical properties in the thermodynamic limit, ($T_c$, $\rho_c$), are then obtained 
by extrapolation of the pseudocritical values using the scaling laws
\cite{Wilding,Perez}.
\begin{equation}
T_c(L) - T_c \propto L^{-(1+\theta)/\nu};
\end{equation}
\begin{equation}
\rho_c(L) - \rho_c \propto L^{-1/\nu}.
\end{equation}
where $\theta$ and $\nu$ are critical exponents.\cite{Blote}

\subsubsection{Gibbs-Duhem Integration}

In order to compute the binodals of the LVE we used a version of Gibbs-Duhem integration (GDI)\cite{kofkeMP} in the GCE \cite{AlmarzaJCP2011,Noya,Hoye}. GDI requires the knowledge of an initial point $(T_0,\mu_0)$ on the phase coexistence line.
Then, the binodals are obtained by solving the differential equation:
\begin{equation}
d \mu =  \left( \frac{\mu}{T} - \frac{\Delta U}{T\Delta N} \right) d T.
\label{gdi}
\end{equation}
The equation is solved numerically using finite intervals, and a fourth-order Runge-Kutta procedure. The $\Delta$'s represent the difference between the mean values of the properties ($U$ or $N$) in the two phases at equilibria, computed by simulations of both phases for the same system size. The sampling of the simulations was enhanced by incorporating cluster moves as described in the Appendix. We used $L=32$ as the system size for the integrations, except at low temperatures where $L$ was increased to $L=64$.

\subsection{Order-disorder transition}

The order-disorder transitions were computed using thermodynamic integration
techniques which give the value of the chemical potential at coexistence, $\mu_0$, at
a given temperature. This can then
be used as the initial coexistence point 
to calculate the binodals using Gibbs-Duhem integration. 

The initial coexistence point is obtained by choosing a value of the chemical potential 
sufficiently high to ensure that the system is fully occupied at $T=0$, 
and computing the equation of state $\rho(\mu_0,T)$ for two sequences of temperatures, one
starting at infinite temperature (disordered phase), and the other starting at $T\simeq 0$
(ordered phase). In these limits the grand potential per site is known exactly. The coexistence 
temperature is determined by the equality of the grand potential of the two phases.

\subsubsection{The ground state}

In the ground state (GS), the stable configurations minimize the grand canonical Hamiltonian: ${\cal H} = U - \mu N$. 
The minimum energy of the model with $N$ occupied sites occurs when all the $A$-patches are
bonded through $AA$ bonds, i.e.  the
potential energy is $U_{GS} =  - N  \epsilon$, which implies:
\begin{equation}
{\cal H}_{GS}(N) = (- \epsilon - \mu ) N;
\label{hgs}
\end{equation}
It follows that the equilibrium state corresponds to the empty lattice
when $\mu < -\epsilon$, and to the full lattice with $M$ $AA$ bonds when $\mu > - \epsilon$. 
The fully occupied ground state is highly degenerate\cite{Confined_SARR} as a large number of configurations are compatible 
with $U_{GS}= - M \epsilon$. In general, the order of a fully occupied configuration may be described through the 
order of a set of planes $[i,j,k]$: Particles in one plane have the same in-plane orientation but 
different planes may exhibit different (in-plane) orientations. There are two types of such planes $[1,1,1]$, 
and $[1,0,0]$. In $[1,1,1]$ planes the sites form a triangular lattice, one site has six NN on the plane, 
and there are three in-plane orientations. 
In $[1,0,0]$ planes the sites form a square lattice, one site has four NN on the plane and there are 
two in-plane orientations.
Taking into account the boundary conditions, the degeneracy of the GS at full occupancy is then: 
\begin{eqnarray}
\Omega_{0} & \simeq  & (3 \times  4^L) +  (4 \times 3^L) 
\label{deggs}
\end{eqnarray}
It follows that the entropy of the ground state scales linearly with $L$ ($S \propto L$) and that the 
entropy per site vanishes in the thermodynamic limit.

The grand canonical partition function $\Xi$ at very low temperatures and at full occupancy 
($\mu > - \epsilon$) is:
\begin{equation}
\Xi_{0}(M,\mu) = \Omega_{0} \exp \left[ + \beta M \epsilon + \beta M \mu \right],
\label{eq_z}
\end{equation}
from where the grand potential can be easily calculated using:\cite{Hansen}
\begin{equation}
\Phi = - k_B T \ln \Xi,
\label{eq_f}
\end{equation}
It follows that at low temperatures in the thermodynamic limit:
\begin{equation}
\phi \equiv \frac{\Phi}{M}  = -\epsilon - \mu;   \;\; ( T \rightarrow 0; \mu > -\epsilon; L  \rightarrow \infty).
\end{equation}
\subsubsection{The high temperature limit}

In the limit of high temperatures, $\beta \epsilon_{\alpha \beta} \rightarrow 0$, the
grand canonical partition function can be written as:
\begin{equation}
\Xi = \sum_{N=0}^{M} \left( \begin{array}{c} M \\ N \end{array}  \right) e^{N \beta \mu } q^N,
\end{equation}
where we took $\beta \mu$ finite.
The factor $q^N$ accounts for the $q$ possible particle orientations. Using elementary combinatorics we find:
\begin{equation}
\Xi = \left( 1 + q e^{\beta \mu} \right) ^M,
\end{equation}
which, as $\beta \rightarrow 0$ at finite $\mu$, yields
\begin{equation}
\ln \Xi = M \ln \left( 1 + q \right).
\end{equation}
Then, the grand canonical potential per site at high temperatures is given by:
\begin{equation} 
\beta \phi =  - \ln (1+q); \;\; (T\rightarrow \infty);
\label{eq_highT}
\end{equation}

\subsubsection{Computing the initial order-disorder coexistence point}

Given the grand canonical potential of one point in each phase, the order-disorder 
transition is located using thermodynamic integration.
In the Grand Canonical ensemble the variation of the thermodynamic potential
at constant volume ($M$) 
is given by:
\begin{equation}
{\textrm d} \left( \beta \phi \right) = 
\frac{U }{M} {\textrm d} \beta - \frac{N}{M}  d \left( \beta \mu \right) =
\frac{\cal H }{M} {\textrm d} \beta - \beta \eta {\textrm d} \mu;
\end{equation}
where $U$ is the potential energy, defined by Eqs. (\ref{u1}-\ref{u2}), and $\eta \equiv N/M$.
This may be used to calculate the thermodynamic potential $\phi(T)$ of the ordered phase as a function of temperature 
by thermodynamic integration at constant $\mu=\mu_0$:
\begin{equation}
d \left( \beta \phi\right) = h (T) d \beta, 
\label{integro_beta}
\end{equation}
where we defined $h\equiv {\cal H}/M$.
In order to calculate the grand canonical potential of the ordered phase Eq. \ref{integro_beta} has to be rewritten 
to avoid divergences at low temperatures. First we add and subtract the value of the integrand at zero temperature and at full
occupancy (which 
is $h_0=-\epsilon-\mu_0$):
\begin{equation}
d( \beta \phi (T)) = h_0  d \beta + \left[ h(T) - h_0  \right] d \beta,
\end{equation}
and then change variable from $\beta$ to $T$. The grand canonical potential of the ordered phase is given by:
\begin{equation}
 \beta \phi(T) = \beta h_0 - \frac{1}{k_B} \int_{0}^T 
\frac{ h(T') - h_0 }{T'^2} d T',
\label{intt0}
\end{equation}
where the integrand is well behaved over the domain of integration, approaching zero as $T \rightarrow 0$.

Similarly, the grand canonical potential $\phi(T)$ of the disordered phase may be computed by integrating Eq. \ref{integro_beta} (using $\beta$ as the integration variable) from the infinite temperature limit at constant volume and $\mu=\mu_0$:
\begin{equation}
\beta \phi(\beta)  = -\ln (1+q) + \int_{0}^{\beta} h(\beta') d \beta',
\end{equation}

\begin{figure}
\includegraphics[width=85mm]{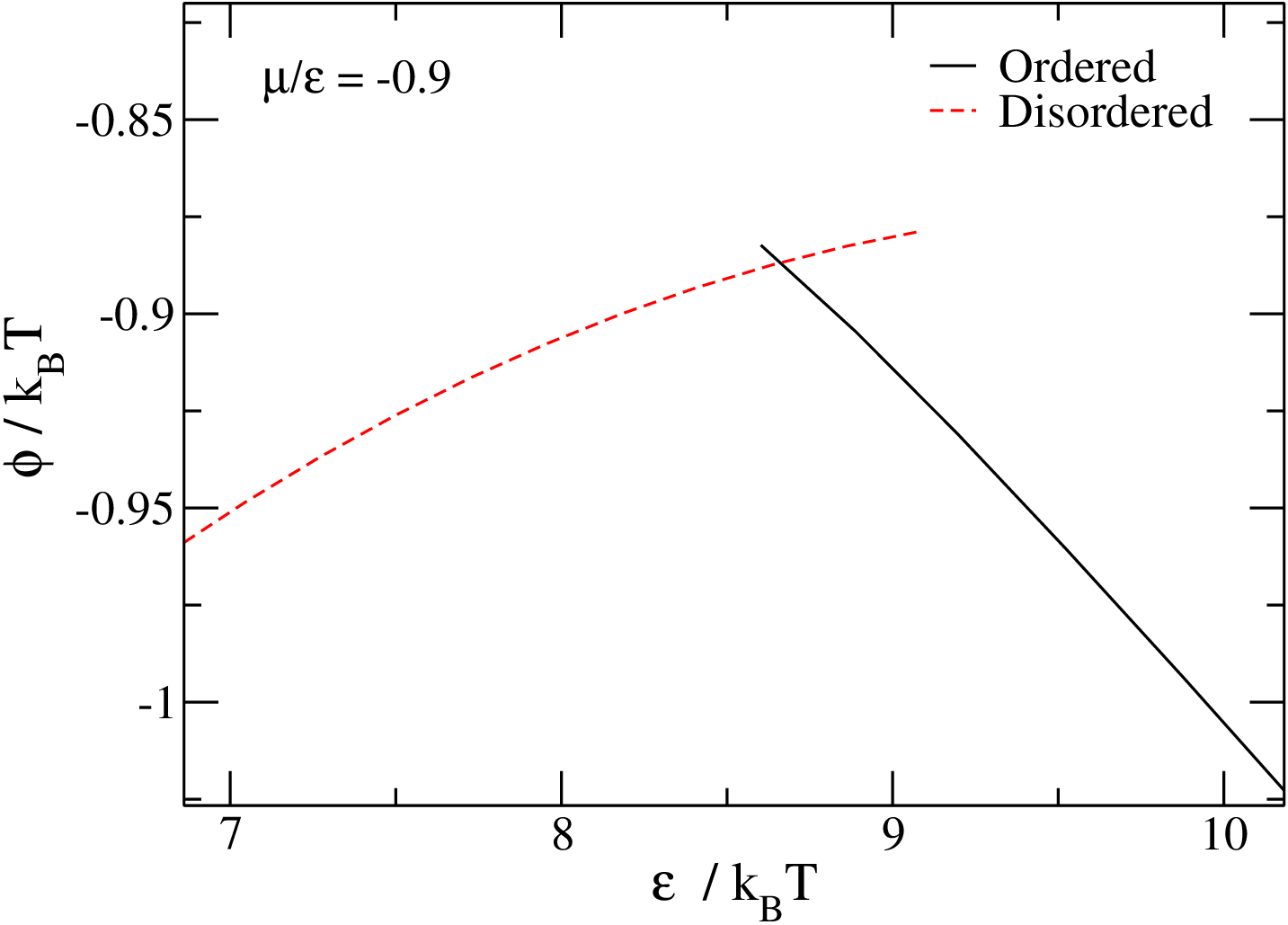}
\caption{Grand canonical potential of the ordered and disordered phases for $r$=0.32
at $\mu/\epsilon = -0.9$. The order-disorder phase transition occurs at the crossing of the two branches.}
\label{fig_order_disorder}
\end{figure}

The order-disorder transition is discontinuous and for sufficiently large systems exhibits hysteresis. This simplifies the calculation of the transition temperature at fixed $\mu$ as it is possible to evaluate ${\cal H}(T)$ for each phase in a range 
of temperatures around coexistence. The two branches of $\Phi(T)$ (ordered and disordered) cross at the transition temperature (see Fig. \ref{fig_order_disorder}). We computed the transition temperatures at $\mu_0/\epsilon=-0.90$ for several values of $r$: 0.00, 0.32, and 0.40. In all cases we found negligible size effects and used $L=24$ , and $L=32$ as typical system sizes.

Having obtained the transition temperature at $\mu=\mu_0$, GDI integration is used to calculate the coexistence 
lines at different temperatures, as was done for the liquid-vapor equilibria. At low temperatures we used large system 
sizes $L=64$, otherwise $L=32$.

\subsection{Consistency checks}

	The calculations of the liquid-vapor and order-disorder transitions
were checked by performing simulations with two fully independent
programs written by two of the coauthors of this paper. In one of the MC codes
cluster sampling techniques were included whereas in the the other (control
program) only simple single site moves were included.  The control code was used
to calculate the liquid-vapor and order-disorder transition using thermodynamic
integration. These calculations validate both the enhanced sampling techniques and the 
proper coding of the Wang-Landau and cluster moves in the GCMC/GDI simulations.

	The calculation of the order-disorder transition using the control
code was carried out using the thermodynamic integration technique described in the previous
section, whereas the calculation of the liquid-vapor equilibria by thermodynamic
integration is described briefly in what follows.

For the vapor phase, the grand canonical potential was obtained integrating
along an isotherm from very low values of the chemical potential (or densities), 
where the system behaves as an ideal gas, to the chemical potential of interest using:
\begin{equation}
\beta \phi (\mu) = \beta \phi(\mu_{ideal}) - \beta \int_{\mu_{ideal}}^{\mu} \eta(\mu') d\mu'
\label{integral_mu}
\end{equation}
where $\mu_{ideal}$ represents a value of the chemical potential low enough so that the behavior
of the fluid can be considered ideal. 
The grand canonical potential of the ideal gas can be easily calculated using:
\begin{equation}
\beta \phi (\mu )= - \beta p 
\end{equation}
and the ideal gas equation to obtain:
\begin{equation}
\beta \phi (\mu_{ideal} )=-\eta
\end{equation}

The free energy of the liquid was calculated through an integration path starting at the high temperature limit, 
where the free energy is calculated as in the previous section (Eq.\ref{eq_highT}).
The integration was performed in two steps:
first, we integrate from infinite temperature to the temperature of interest 
keeping the chemical potential $\mu $ constant  (Eq. \ref{integro_beta})
and, second, we integrate along an isotherm from $\mu $ to the chemical potential of interest
(Eq. \ref{integral_mu}). As usual we ensure that there are no phase transitions along the 
chosen thermodynamic path.

	Analogously to the calculation of the order-disorder transition the liquid-vapor coexistence is obtained by calculating, 
at the given temperature, the chemical potential where the grand canonical potentials of the two phases are equal.

	All the simulations performed with the control Monte Carlo code considered systems with $L=$24. The results obtained using the two different codes and methodologies 
are found to be consistent.
	
	
\section{Results}
\label{sec:Results}

\begin{figure}
\includegraphics[width=150mm,clip=]{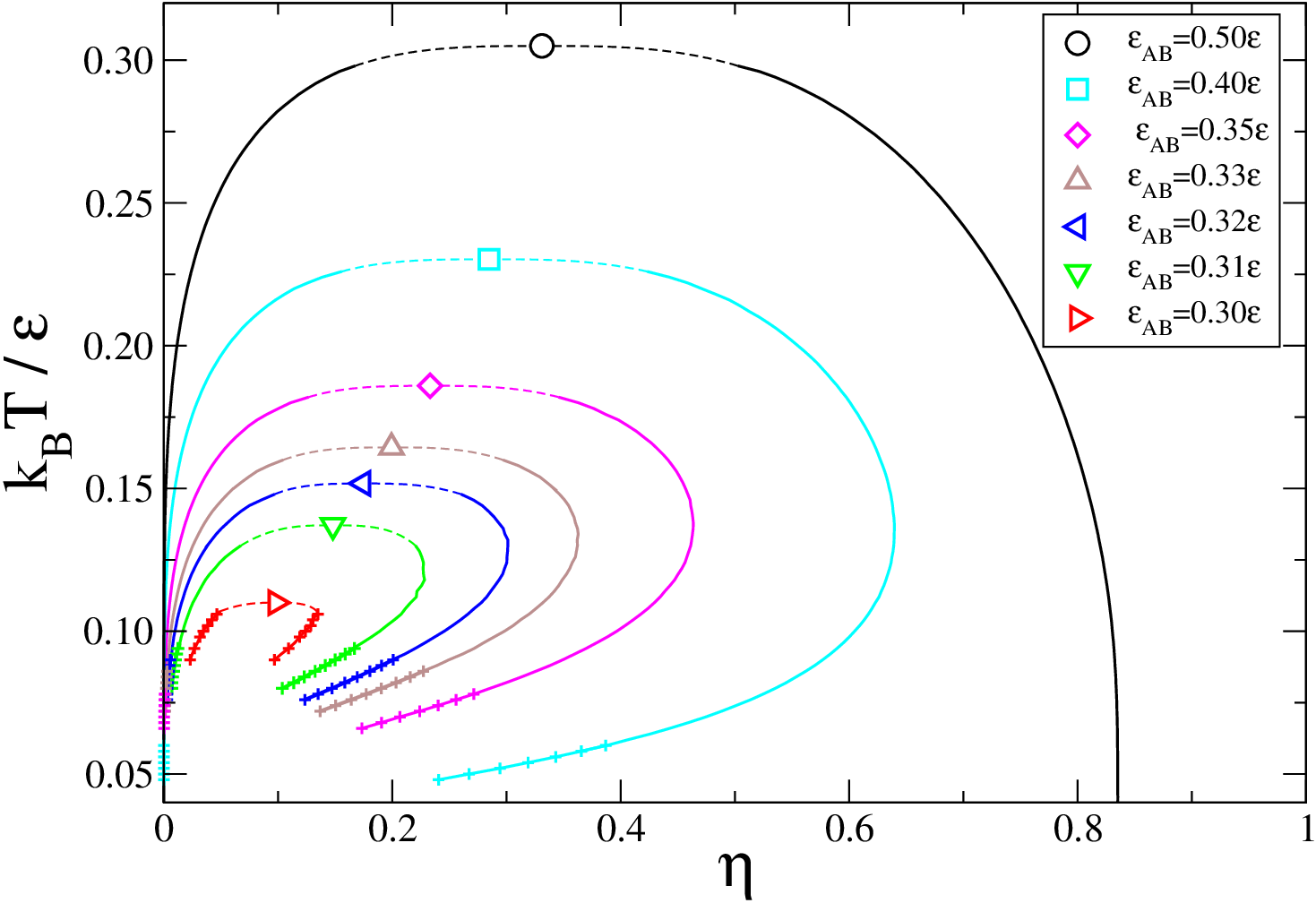}
\caption{Liquid-vapor binodals, for different $r=\epsilon_{AB}/\epsilon$. Large open
symbols mark the critical points. Continuous lines represent GDI results, the points on these
lines mark the portions that were  computed using larger system sizes ($L=64)$.
Dashed lines are included to connect the GDI results with the critical point estimates.}
\label{Fig01}
\end{figure}

In Figure \ref{Fig01} we plot the results for the liquid-vapor equilibria, including the estimates for the critical point (these are also given in Table \ref{tcs}). 
The general trend is as expected from earlier work\cite{Russo1,Russo2,AlmarzaJCP2011}: As $r$ decreases
the critical point is shifted to lower densities and lower temperatures. In addition for $r < 1/2$ the LVE 
is reentrant, with densities of the liquid phase at coexistence decreasing on cooling at low temperatures. 
The LVE binodals were computed using GDI with system size $L=32$ (i.e. $M=$ 131 072) for $r>0.30$ (except at 
the lowest temperatures, where $L=64$ ($M=$ 1 048 576) was used to avoid interconversion between the two phases). For $r=0.30$ larger systems, $L=64$, were required to obtain reliable results at all temperatures. In all cases, 
the GDI fails at sufficiently low temperatures, due to the rapid growth of the typical size of the voids in the (emptying) coexisting liquid as reported previously \cite{AlmarzaJCP2011}. 
As in other $2AnB$ models \cite{Russo1,Russo2,AlmarzaJCP2011,Almarza_closedloop} the binodal at a given $r$ encloses 
the binodals at smaller values of $r$.

The most remarkable result, however, is the clear evidence of LVE for systems with $r < \frac{1}{3}$. As mentioned in the Introduction, earlier theoretical predictions based on Wertheim's first-order perturbation theory set the threshold for liquid-vapor coexistence at $r=\frac{1}{3}$ in line with simulation results for $2AnB$ models \cite{Russo2,AlmarzaJCP2011}. The finding is also relevant in a wider context, as this threshold was used recently to address the liquid-vapor condensation in other systems that form branched chains at low temperatures, in particular dipolar hard spheres\cite{TavaresMP2011}. We will return to this discussion later, after the theoretical analysis described in Sec. \ref{sec:theory}.

\begin{figure}
\includegraphics[width=150mm,clip=]{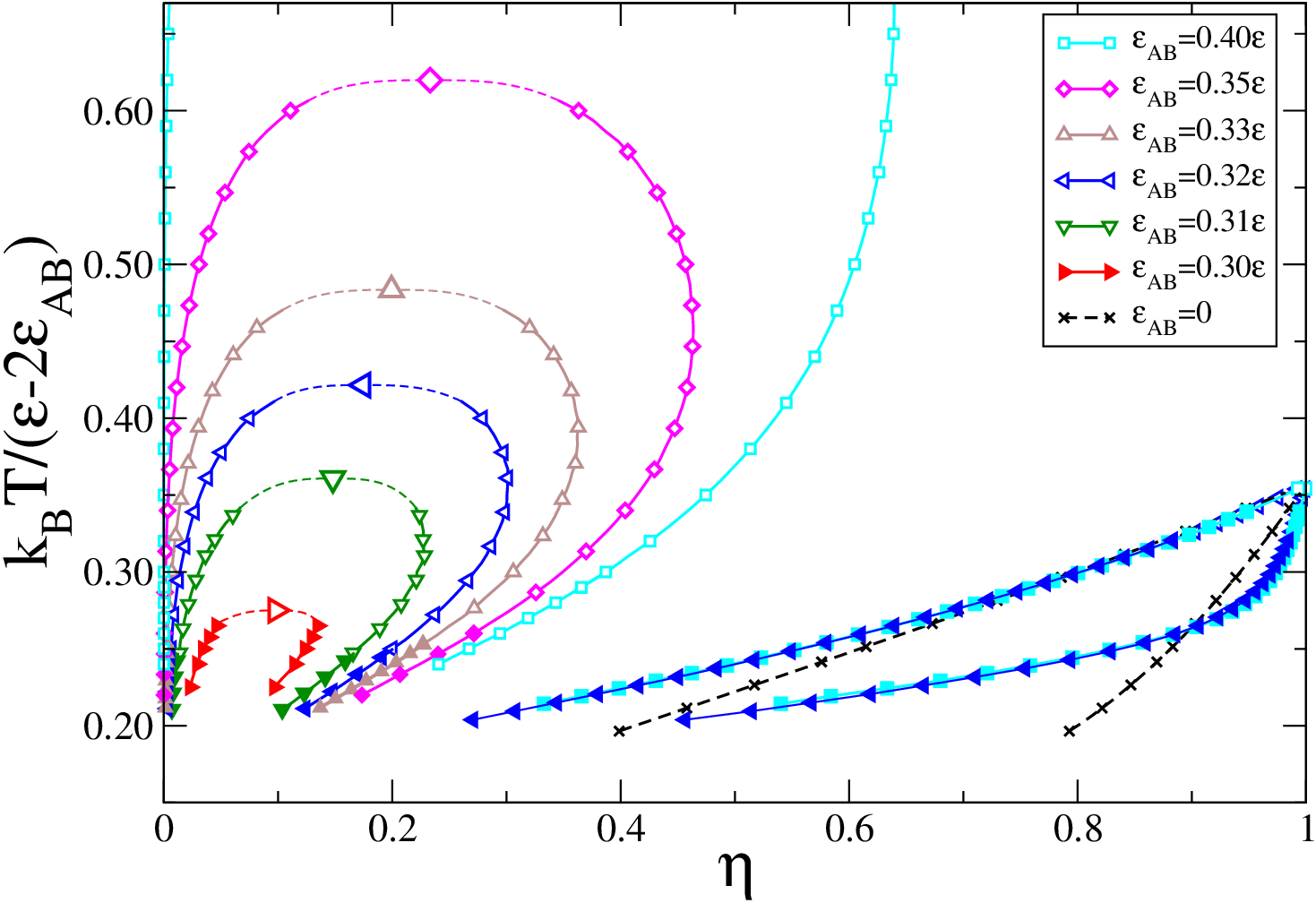}
\caption{Binodals for the liquid-vapor and order-disorder transitions, for different $r=\epsilon_{AB}/\epsilon$, as functions of a rescaled temperature. Filled symbols mark
the portions of the curves that were computed using larger system sizes (L=64).}
\label{Fig02}
\end{figure}

Lattice models allow the precise location of not only the liquid-vapor transition, but also the order-disorder 
transition that occurs at higher densities (the lattice analogue of the fluid-solid transition of off-lattice models). As in  the 2D lattice, we find that the order-disorder transition occurs always at a higher density 
(higher chemical potential), in the temperature range accessible to simulations (see Fig. \ref{Fig02}) \cite{AlmarzaJCP2011}. 

Interesting scaling features are revealed by plotting the LVE and the order-disorder binodals for different values of $r$ as functions of the scaled temperature, $t=k_BT/[(1-2r)\epsilon]$
(see Fig. \ref{Fig02}).
First, in the limit of full occupancy the order-disorder transitions collapse into a single point. This is an exact result, and the observed collapse is just a check of the consistency and accuracy of our simulation protocols. In addition, the order-disorder transition exhibits a very weak dependence on $r$; i.e. the lines for the order-disorder transition for $r=0.32$, and $r=0.40$ are almost indistinguishable. By contrast, the order-disorder transition of the SARR model ($r=0$), deviates clearly from the previous ones.
Finally, as in the T-$\eta$ representation, the binodal for a given $r$ encloses the binodals for lower values of $r$. 

\begin{table}
\begin{tabular}{c|lll}
\hline
$r$   &    $T_c^*$    &  $\eta_c$   &  $\mu_c$   \\
\hline
0.300 &  $0.1163(9) $   & $0.105(4)$ & -1.0361(10)\\
0.305 & 0.1277(3)  & 0.1284(10)  &  -1.0522(4)  \\
0.310 &  0.1371(4)  &  0.148(2)  & -1.0679(6) \\
0.320 & 0.1518(2) & 0.175(2) & -1.0961(3) \\
0.330 & 0.1644(1) & 0.1993(10) & -1.1245(2) \\
0.350 & 0.1860(1) & 0.2331(2) & -1.1808(2) \\
0.400 & 0.2302(1) & 0.2848(3) & -1.3228(1) \\
0.450 & 0.2689(1) & 0.3136(3) & -1.4684(1) \\
0.500 & 0.3050(1) & 0.3310(2) & -1.6169(1) \\
\hline \hline
\end{tabular}
\caption{Estimates of the critical points for different values of $r$}
\label{tcs}
\end{table}

\section{Wertheim's theory for $2AnB$ lattice models}
\label{sec:theory}
The free energy per particle, within Wertheim's first order perturbation theory, for a homogeneous system of particles 
with 2 patches of type $A$ and $n$ patches of type $B$ is \cite{Tavares,Tavares_MP09},
\begin{equation}
\label{fenWerth}
\beta f = \beta f_{ref}+ 2(\ln X_A -\frac{X_A}{2}) + 
n (\ln X_B -\frac{X_B}{2}),
\end{equation}
where $f_{ref}$ is the free energy per particle 
of the reference system and $X_\alpha$ the 
fraction of unbonded patches of type $\alpha$.
The laws of mass action that relate $X_\alpha$, the density $\eta$ and the 
temperature $T$ are (when there are no $BB$ interactions)\cite{Tavares,Tavares_MP09},
\begin{equation}
\label{lma1}
X_A+2\eta\Delta_{AA}X_A^2+n\eta\Delta_{AB}X_AX_B=1,
\end{equation}
\begin{equation}
\label{lma2}
X_B+2\eta\Delta_{AB}X_AX_B=1.                        
\end{equation}
As for the $2A2B$ model on the square lattice,   
the reference 
system is an ideal lattice gas \cite{AlmarzaJCP2011}, and thus,
\begin{equation}
\beta f_{ref}= \ln \eta +\frac{1-\eta}{\eta}\ln(1-\eta).
\end{equation}
The  $\Delta_{\alpha\beta}$ are integrals of the Mayer functions of 
two patches $\alpha$ and $\beta$ on two different particles, 
over their positions and orientations,  
weighted by the pair distribution function of the reference system. In 
continuous systems the $\Delta_{\alpha\beta}$ are calculated from,
\begin{equation}
\label{DABcont}
\Delta_{\alpha\beta}=\frac{1}{V(4\pi)^2}\int d\vec r_1 d\vec r_2 
\int d\vec \omega_1 d\vec \omega_2 g_{ref}(\vec r_1, \vec r_2)
f_{\alpha \beta}(\vec r_1,\vec r_2, \vec \omega_1,\vec \omega_2),
\end{equation}
where $\alpha$ is a particular patch on particle 1 and $\beta$ is a particular
patch on particle 2, $\vec r_i$ refers to the position of particle 
$i$ and $\vec \omega_i$ to the orientation of the particular patch on particle 
$i$ that is being considered; the factor $1/(4\pi)^2$ takes into account that 
there is no preferred orientation for the position of the patches on the particles surfaces. 
Finally, $f_{\alpha\beta}(\vec r_1,\vec r_2, \vec \omega_1, \vec \omega_2)$ is the Mayer function of the interaction potential between patches $\alpha$ and $\beta$, and $g_{ref}(\vec r_1, \vec r_2)$ is the pair correlation function of the reference system\cite{Tavares,Tavares_MP09}.

The calculation of $\Delta_{\alpha\beta}$ on a lattice with coordination number $z$ and particles with $z$ patches, 
(as in the $2AnB$ models considered here and in \cite{AlmarzaJCP2011}) is carried out by discretizing Eq. \ref{DABcont}, 
\begin{equation}
\label{DABlatt}
\Delta_{\alpha\beta}=\frac{1}{M z^2}
\sum_{i_1=1}^{M} \sum_{i_2=1}^{M}
\sum_{l_1=1}^z \sum_{l_2=1}^z f_{\alpha\beta}(i_1,i_2,l_1,l_2) g_{ref}(i_1,i_2).
\end{equation}
Here the patch $\alpha$ is on particle $1$ and the patch $\beta$ on particle 
$2$; $i_1$ and $i_2$ represent the positions of particles 1 and 2, respectively; 
$M$ is the total number of lattice sites (equivalent to the volume); the factor $1/z^2$ accounts for the fact that 
there is no preferred orientation for the patches, as $z$ is also the number of different orientations of a given patch.
The integers $l_1$ and $l_2$ run over the possible orientations of each patch. For the potential described in Sec. \ref{sec:3DModel}, the Mayer function $f_{\alpha\beta}(i_1,i_2,l_1,l_2)$ is non zero when:     
 (a) $1$ and $2$ are NN and (b) $l_1$ and $l_2$ are such that patches 
 $\alpha$ and $\beta$  are 
properly oriented along the interparticle direction.  Using $g_{ref}=1$, Eq. \ref{DABlatt} is simplified to,
\begin{equation}
\label{DABlatt2}
\Delta_{\alpha\beta}=v_{\alpha\beta}\left[\exp(\beta\epsilon_{\alpha\beta})-1\right],
\end{equation}
where $v_{\alpha\beta}$, the volume of a bond between 
patches of type $\alpha$ and $\beta$, is 
\begin{equation}
\label{vab}
v_{\alpha\beta}=1/z.
\end{equation}
For the $2AnB$ models under consideration, the bonding volume $v_{\alpha\beta}$ is independent of the types of patches and 
is related to the number of $B$ patches $z=n+2$. 

\begin{figure} 
\includegraphics[width=150mm,clip=]{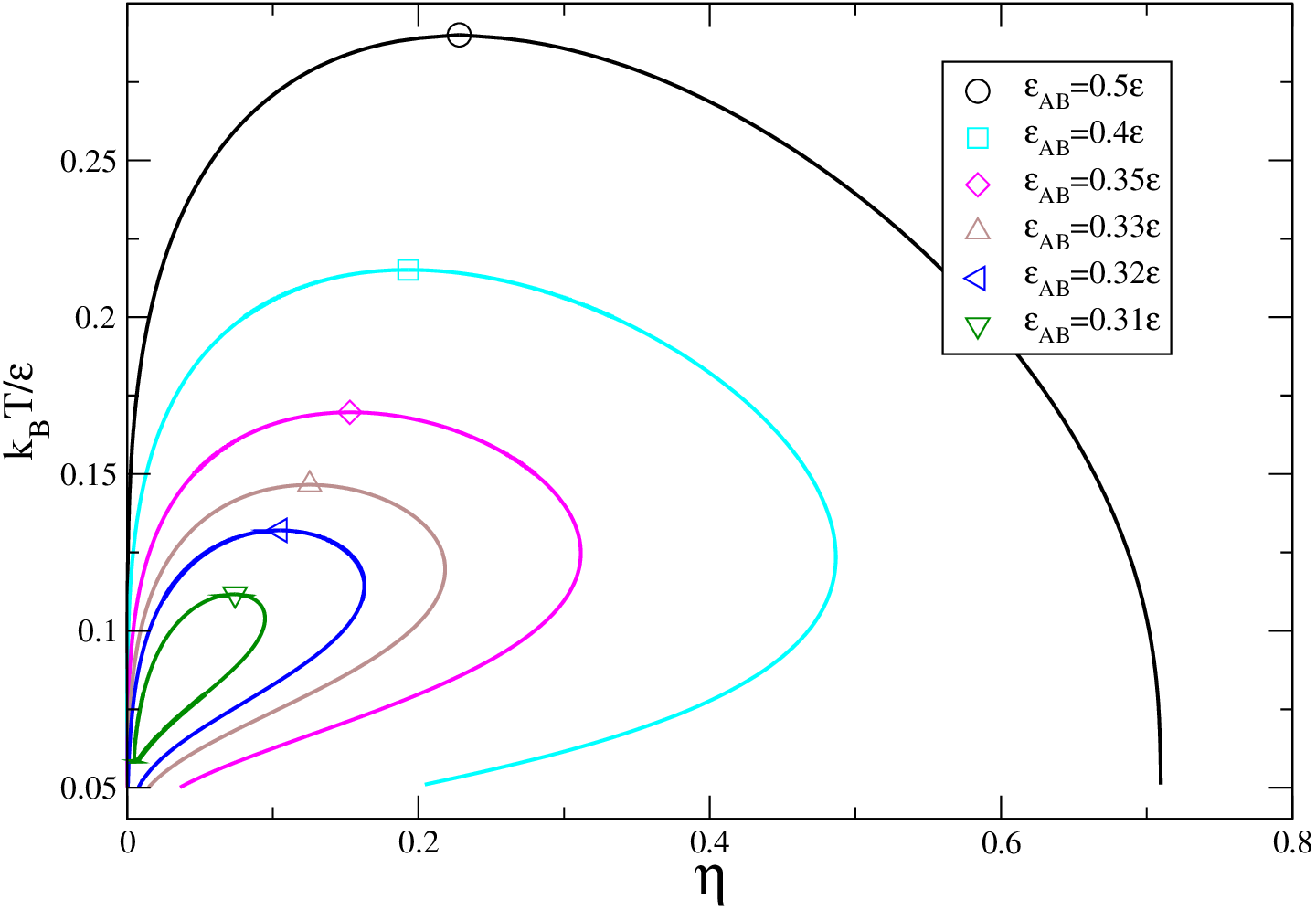}
\caption{Liquid-vapor binodals, for different $r=\epsilon_{AB}/\epsilon$, calculated using Wertheim's first order perturbation theory. Larger critical temperatures correspond to larger values of $r$.}
\label{fig:phdfB10}
\end{figure}

The liquid-vapor binodals calculated using Wertheim's theory, for models with $n=10$, i.e. the models simulated in Sec. \ref{sec:Results}, are plotted in Figure \ref{fig:phdfB10}, for several $r$. 
As expected \cite{AlmarzaJCP2011,Russo2,Russo1}, Wertheim's theory predicts, in agreement with the simulation results, the reentrance of the liquid binodal for $r<\frac{1}{2}$. The phase diagrams were calculated for models with  $r > 0.305$: below this value no liquid-vapor coexistence was found.

\begin{figure} 
\includegraphics[width=150mm,clip=]{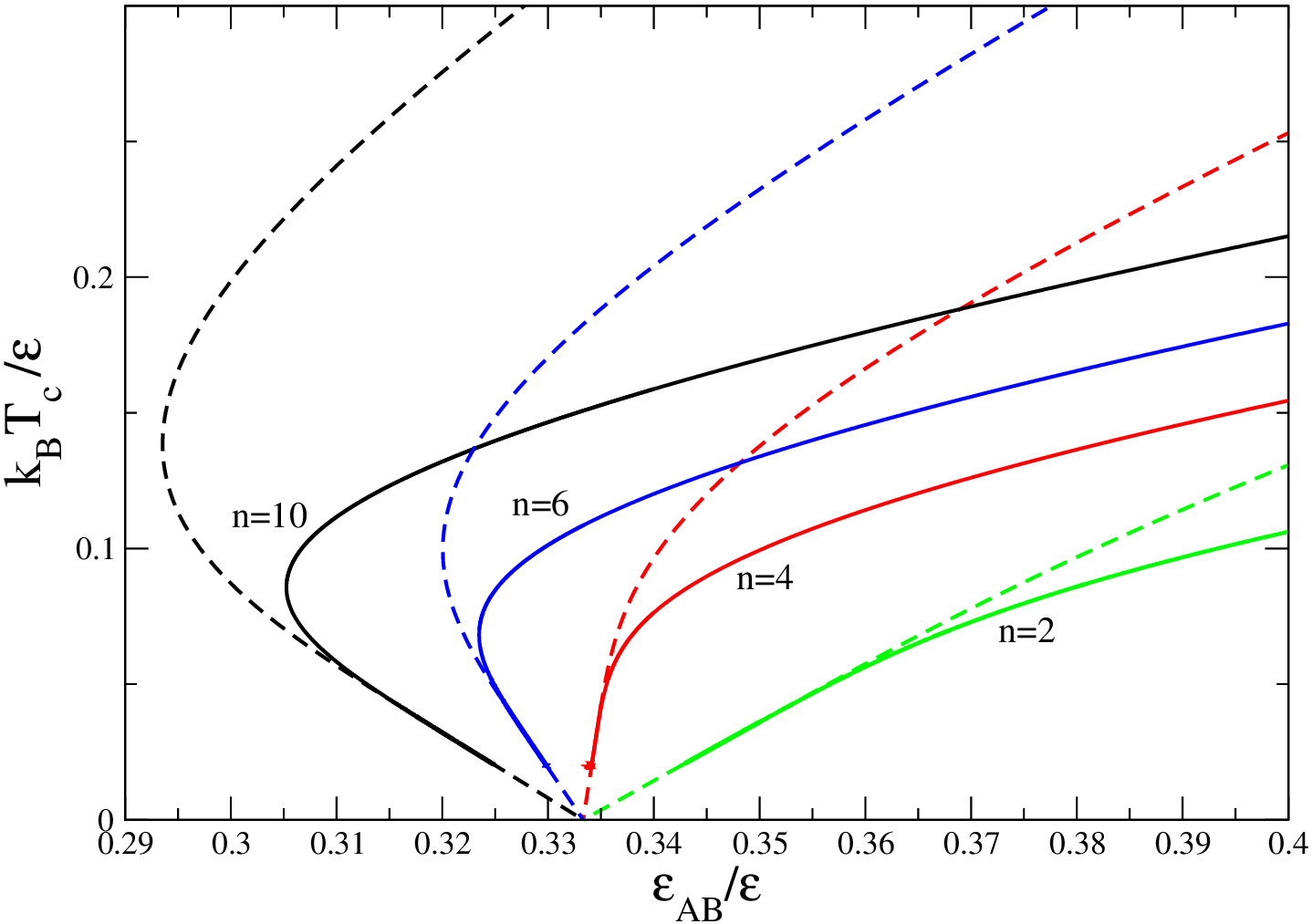}
\caption{
Critical temperature as a function of $r\equiv \epsilon_{AB}/\epsilon$, for several values
of $n$. Full lines: critical temperature calculated from Wertheim's  
theory Eq. \ref{fenWerth}; dashed lines: critical temperature calculated 
from the asymptotic expansion of Wertheim's theory, Eqs. (\ref{Tcasympt}) and (\ref{Cconst}).}
\label{TceAB}
\end{figure}

\begin{figure} 
\includegraphics[width=150mm,clip=]{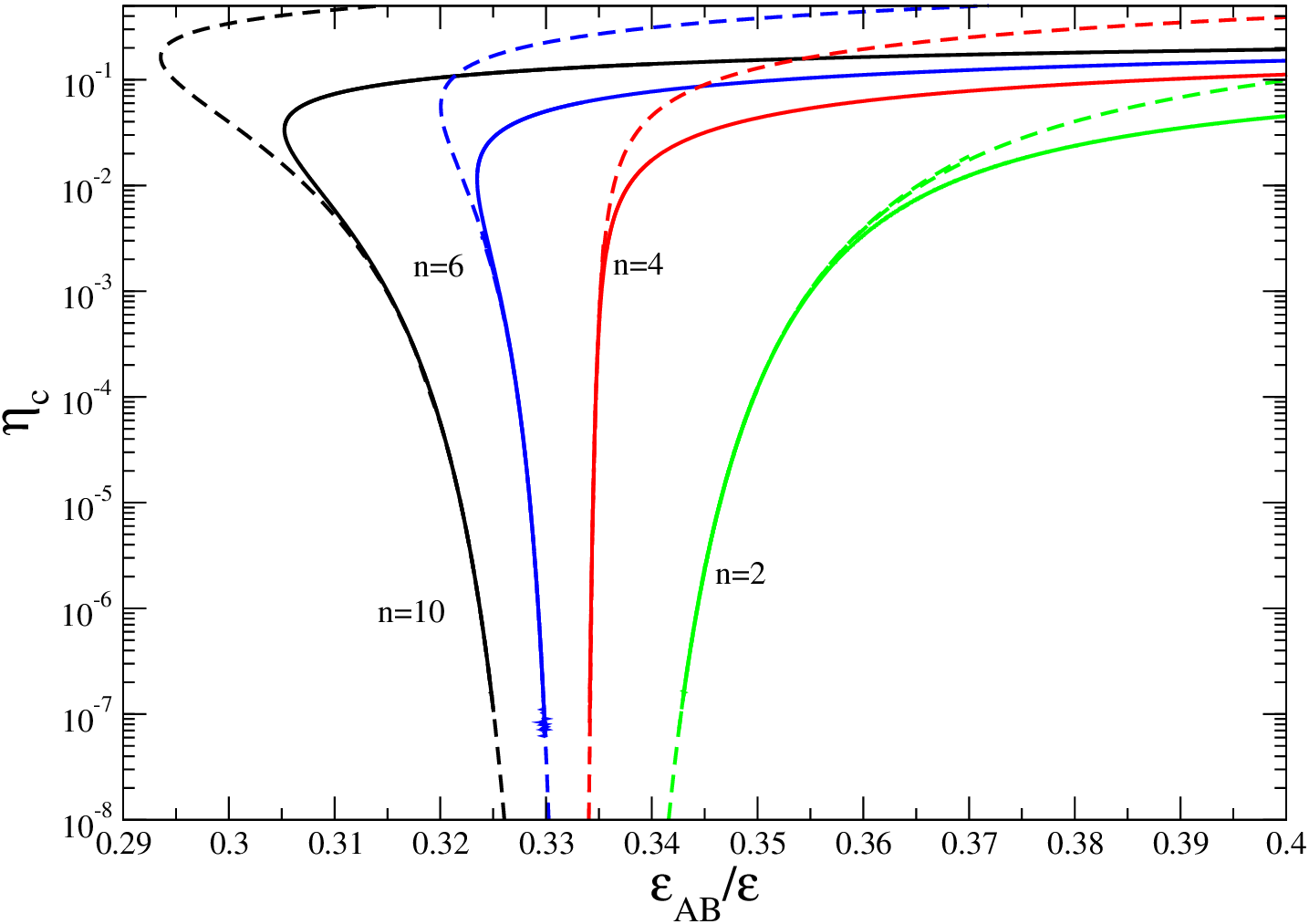}
\caption{
Critical density as a function of $r$ for several values
of $n$. Full lines: critical density calculated from Wertheim's 
theory Eq. \ref{fenWerth}; dashed lines: critical temperature calculated 
from the asymptotic expansion of Wertheim's theory, Eqs. (\ref{Tcasympt}), (\ref{Cconst})
and (\ref{rhoc}).}
\label{rhoceAB}
\end{figure}

To clarify this surprising behavior, in view of previous results \cite{Tavares,Tavares_MP09,Russo2,Russo1,AlmarzaJCP2011}, we solved the set of equations, $\left(\frac{\partial^2 f}{\partial\eta^2}\right)_T=0$ and $\left(\frac{\partial^3 f}{\partial\eta^3}\right)_T=0$,  which determine the critical density and temperature as a function of the parameters of the lattice $2AnB$ model, namely $r$ and $n$ (or the coordination number $z$). The results for $n=2$, $n=4$, $n=6$ and $n=10$ (corresponding to square \cite{AlmarzaJCP2011}, triangular or simple cubic, body centered cubic, and face centered cubic lattices, respectively) are plotted in Figs. \ref{TceAB} and \ref{rhoceAB}. 


The results reveal that Wertheim's theory predicts a phase behavior that is strongly dependent on the values of $n$ 
and $r$. For $n < 4$ we recover the threshold reported earlier: no critical point for $r< \frac{1}{3}$ and one critical point otherwise. For larger $n$, however, three distinct regimes are possible depending on 
the value of $r$: one critical point for $r > \frac{1}{3}$, no critical point for $r$ less than a threshold $r_{m}$ ($<\frac{1}{3}$) and two critical points for the range of $r$ between this threshold and $\frac{1}{3}$.

A deeper understanding of these results, which contrast with the simpler picture reported earlier \cite{Tavares,Tavares_MP09,Russo2,Russo1,AlmarzaJCP2011}, is obtained through the asymptotic expansion of the free energy 
Eq. \ref{fenWerth}, in the limit of strong $AA$ bonding (i.e $X_A\approx 0$) and low densities  \cite{Tavares,Tavares_MP09,Russo2,Russo1}. Using this expansion we find for the pressure $p$,
\begin{equation}
\label{presasympt}
\beta p = \frac{\eta^{\frac{1}{2}}}{\sqrt{2\Delta_{AA}}}-
\frac{n\Delta_{AB}}{\sqrt{2\Delta_{AA}}}\eta^{\frac{3}{2}}+
B_2\eta^2,
\end{equation}
where $B_2$ is the second virial coefficient of the reference system ($B_2=1/2$ for the ideal lattice gas). Using Eq. \ref{presasympt} the critical temperature $T_c$ is found to satisfy,
\begin{equation}
\label{Tcasympt}
G(\epsilon_{AB},T_c)\equiv
\frac{\left[\exp\left(\frac{\epsilon_{AB}}{k_BT_c}\right)-1\right]^3}
{\exp\left(\frac{\epsilon}{k_BT_c}\right)-1}=C,
\end{equation}
where $C=\frac{8 B_2^2 v_{AA}}{(n v_{AB})^3}$. For the $2AnB$ models under consideration, the constant $C$ 
may be evaluated using Eq. \ref{vab} and $B_2=1/2$, 
\begin{equation}
\label{Cconst}
C=\frac{2(n+2)^2}{n^3}.
\end{equation}
The asymptotic critical density $\eta_c$ is then given by,
\begin{equation}
\label{rhoc}
\eta_c=\frac{1}{n\Delta_{AB,c}},
\end{equation}
where $\Delta_{AB,c}$ is the value of $\Delta_{AB}$ at $T=T_c$.
In Figures \ref{TceAB} and \ref{rhoceAB} the asymptotic critical temperature and density are plotted as functions of $r$ for several values of $n$. As expected, the asymptotic results describe those of the full theory at low temperatures, but qualitative agreement is obtained at intermediate temperatures. Thus, the phase behavior of the $2AnB$ model can be described by analyzing Eq. \ref{Tcasympt}.

For $r>\frac{1}{3}$ the function $G(\epsilon_{AB},T_c)$ is a monotonic decreasing function of $T_c$, with limits $\lim_{T_c\to 0}=\infty$ and $\lim_{T_c\to \infty}=0$. However, for $r<\frac{1}{3}$, this function
has a maximum which decreases with decreasing $r$; is limited by $0<G(\epsilon_{AB},T_c)<1$ and vanishes in the limits, $\lim_{T_c\to 0}=0$ and $\lim_{T_c\to \infty}=0$. 
The analysis of the critical behavior of $2AnB$ models is done most simply by considering the cases $C>1$ and $C<1$ 
(which correspond, Eq.\ref{Tcasympt}, to $n\le 4$ and $n>4$, respectively, for integer values of $n$).  
\begin{itemize}
\item[$C>1$:]{

If $r>\frac{1}{3}$, Eq. \ref{Tcasympt} has a unique solution, 
and therefore, there is a single critical point with critical temperature,
\begin{equation}
\label{Tclin}
T_c\approx\frac{3r-1}{\ln C}.
\end{equation}
If $r<\frac{1}{3}$ then $G(\epsilon_{AB},T_c)<1$ and Eq. \ref{Tcasympt} has no solution. Therefore, there is no critical point.

This is the case considered in previous work \cite{Tavares,Tavares_MP09,Russo2,Russo1}, as the models analyzed previously have $C>1$.
}
\item[$C<1$:]{
If $r>\frac{1}{3}$, Eq. \ref{Tcasympt} has one solution and there is a single critical point. 
If $r<\frac{1}{3}$, we define $r_{m}$ as the value of $r$ where the maximum of $G$, $G_{max}$, is equal to $C$: $G_{max}(r_{m},T_c)=C$. Then, two cases have to be distinguished:
If $r_{m}<r<\frac{1}{3}$, $G_{max}(\epsilon_{AB},T_c)> C$, 
and Eq. \ref{Tcasympt} has two solutions; therefore, there are two critical points;
If $r<r_{m}$, Eq. \ref{Tcasympt} has no solutions, and there is no critical point. 
} 
\end{itemize}

\begin{figure} 
\includegraphics[width=150mm,clip=]{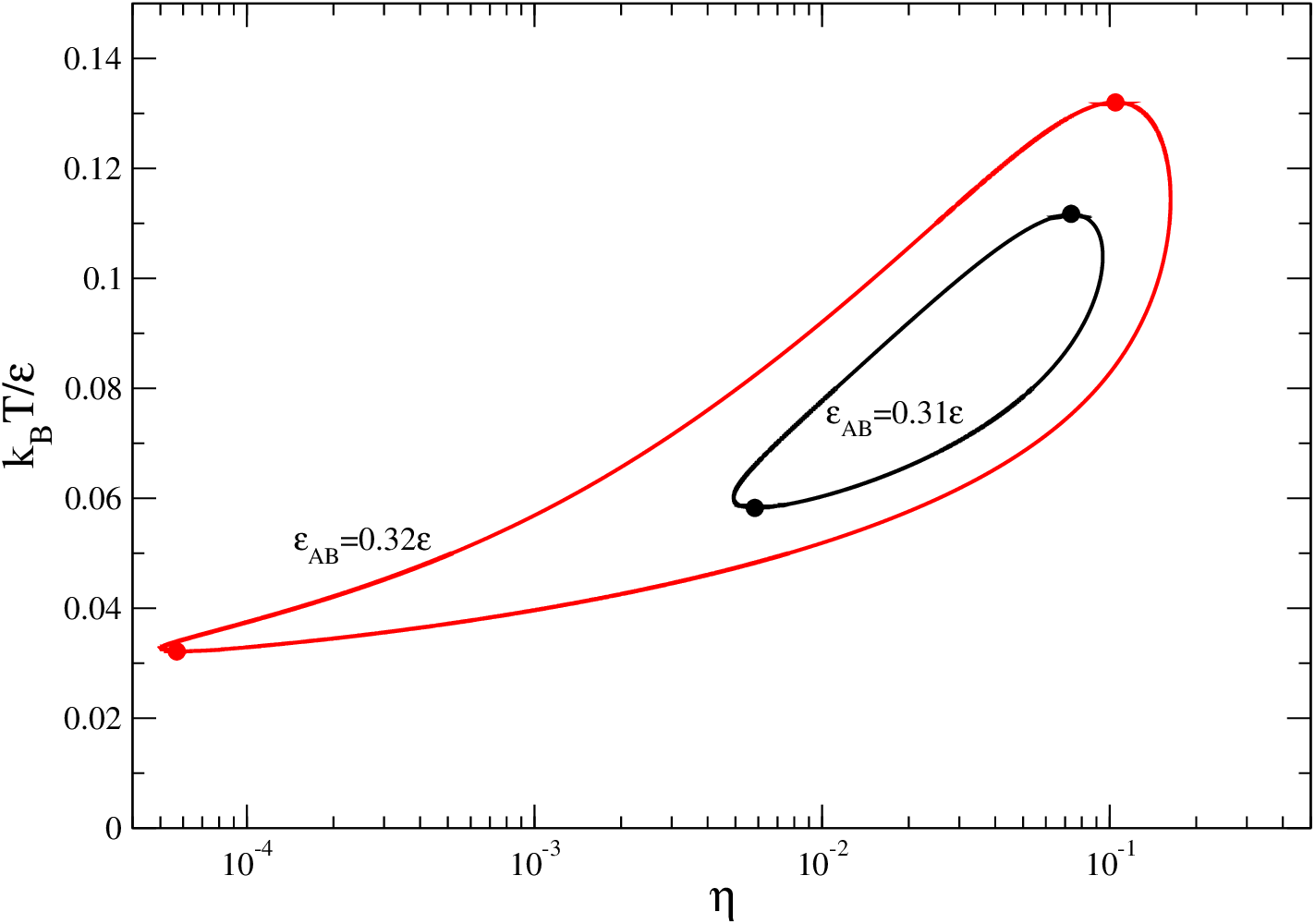}
\caption{
Phase diagrams with two critical points, for $n=10$ and two values of 
$r<1/3$, calculated using Wertheim's theory Eq. \ref{fenWerth}. The circles represent the location of the 
critical points. These binodals are the same as those of figure \ref{fig:phdfB10} 
for the indicated values of 
$r=\epsilon_{AB}/\epsilon$, but are represented in a logarithmic scale in the density and extended to lower temperatures, to put highlight  the lower critical points.}
\label{2critpoint}
\end{figure}

In Figure \ref{2critpoint} we plot a phase diagram with two critical points, for models with $n=10$, and $r=0.31$ and $0.32$.

The liquid vapor coexistence of these models is between a low density, high energy and high entropy phase, formed by short 
chains, and a high density, low energy and low entropy phase (network liquid) formed by long chains connected by $AB$ bonds, 
or junctions \cite{Russo1}. It has been shown \cite{Russo1} that this coexistence is only possible when a decrease in the entropy of chains (or $AA$ bonds) upon condensation, is balanced by an increase in the entropy associated with the junctions (or $AB$ bonds). For systems with $C>1$ and $r<\frac{1}{3}$, the increase in the entropy of the junctions is no longer sufficient to balance the loss in the entropy of the chains \cite{Russo1}. 

Systems with $C<1$ have not been discussed earlier as the continuum \cite{Russo1,Tavares} and the lattice \cite{AlmarzaJCP2011} models investigated previously belong to the class $C>1$. The $2A10B$ lattice model investigated in this paper belongs to the class $C<1$ and thus exhibits different critical behavior. In these models the balance of entropies may occur at values of $r<\frac{1}{3}$. This may be rationalized by considering the physical meaning of $C$. The entropy of one bond is the logarithm of the volume available (on one particle) to form that bond \cite{Douglas2007}. Then, $\ln C= \ln(\frac{2v_{AA}}{2B_2})-3\ln(\frac{nv_{AB}}{2B_2})$ is the difference between the entropy of one $AA$ bond and (three times) the entropy of one $AB$ bond. For the $2AnB$ model $\ln C= \ln(\frac{2}{n+2})-3\ln(\frac{n}{n+2})$, and as $n$ increases so does the entropy of the $AB$ bonds. Therefore, when $C<1$ the entropy of junction formation increases, in such way that it can balance the decrease of entropy of the chains, for values of $r<\frac{1}{3}$.

\section{Discussion of the results}
\label{sec:Discussion}

Despite the challenges posed by the simulations of the phase diagram of empty
fluids at low temperatures, the results for the $2A10B$ lattice model confirm the
features of the LVE reported for 3D off-lattice \cite{Russo2,Russo1} and 2D lattice
$2AnB$ \cite{AlmarzaJCP2011} models. The variation of the critical densities and
temperatures with $r$ follow the expected behavior \cite{Russo2,Russo1,AlmarzaJCP2011}.
In addition, we computed the order-disorder transition that occurs
at higher densities, confirming that the low density liquid phase is
thermodynamically stable as in the 2D model \cite{AlmarzaJCP2011}.

We have, however, found an unexpected result: LVE for systems with
$r<\frac{1}{3}$, by contrast to previous simulation results and the
theoretical analyses based on Wertheim's theory \cite{Tavares,Tavares_MP09,Russo2,Russo1} as
well as an earlier prediction based on a hierarchical theory of network fluids
\cite{Tulsty2000}. The threshold $r=\frac{1}{3}$ results from an asymptotic expansion of 
Wertheim's first-order perturbation theory, which assumes that the constant $\ln C$ is positive as 
described in Sec. \ref{sec:theory}. This is certainly the case for $2AnB$ models on and
off-lattice if the number of $B$ patches is not too large. For lattice models, however, the bonding volume 
compatible with a single bond per patch assumed by Wertheim's theory is much larger than in similar off-lattice 
models (the exclusion of the second particle being guaranteed by the lattice structure) and thus $\ln C$ can become 
negative. In this case, Wertheim's first-order perturbation theory and its asymptotic expansion in the limit of strong 
$AA$ bonding predicts indeed the possibility of LVE for $r<\frac{1}{3}$. The theoretical analysis also
predicts that in this regime the reentrancy of the liquid-vapor binodal is extreme in the sense that the system exhibits a low temperature critical point. The theoretical prediction is then that when $\ln C$ is negative (large values of $n v_{AB}$) $2AnB$ models exhibit a closed miscibility loop in a range of $r<\frac{1}{3}$. There is also a new threshold, which depends on the number of $B$ patches, below which the closed miscibility loop vanishes and where there is no condensation.   

Previous simulation results on and off lattice were compatible with the original $\frac{1}{3}$ threshold,
and in line with the theoretical results for positive $\ln C$ \cite{Russo1,Russo2,AlmarzaJCP2011}. 

The closed miscibility loop predicted for systems with negative $\ln C$ has not been confirmed by
simulations, since the density and temperatures at which they occur are beyond
the current simulation techniques. 

In related work, a $2A4B$ 2D lattice model with $\frac{1}{2} > r > \frac{1}{3}$ was shown to exhibit
the usual reentrant behavior when the position of the $A$ patches prevents
the formation of rings, a closed miscibility loop when the orientation of the
$A$ patches promotes relatively large rings and no phase coexistence when the
orientation of the $A$ patches promotes short rings \cite{Almarza_closedloop}. The topology
of the phase diagram of this $2B4A$ lattice model with $r > \frac{1}{3}$
changes as the orientation of the $A$ patches changes (promoting the formation of rings) in a fashion 
that resembles the behavior of the $2A10B$ model as $r$ decreases. Although the physics may be related 
a detailed, quantitative and qualitative, analysis is required in order to investigate the  analogies in the driving mechanisms of 
the different transitions.

Along these lines, recent work for a model with $A$ patches addressed quantitavely the competition between ring and chain formation, within an extension of Wertheim's first-order perturbation theory and by simulation \cite{Tavares2012}.     
An extension of this approach to $2AnB$ models and the calculation of the corresponding phase diagrams is a challenging task that will be addressed in future work. Likewise new simulation algorithms will be developed to confirm
the presence of closed miscibility loops, in systems with no rings, as predicted by Wertheim's first-order perturbation theory as well as the new thresholds, $r < \frac{1}{3} $, for models with negative $\ln C$. 

The degree of universality of the new thresholds is also an important open question, in general, and in the context of the condensation of dipolar hard-spheres.

\acknowledgements  

NGA and EGN gratefully acknowledge financial support from the Direcci\'on
General de Investigaci\'on Cient\'{\i}fica  y T\'ecnica under Grant No.
FIS2010-15502, from
the Direcci\'on General de Universidades e Investigaci\'on de la Comunidad de Madrid under Grant
No. S2009/ESP-1691 and Program MODELICO-CM.

MMTG, JMT and NGA acknowledge 
financial support from the Portuguese Foundation for Science and Technology 
(FCT) under Contracts nos. PEst-OE/FIS/UI0618/2011 and PTDC/FIS/098254/2008.

\section{Appendix: Cluster algorithms}
In this appendix the cluster algorithms that we used in the Grand Canonical 
ensemble simulations of the GDI method are described. The two cluster moves defined here
do not include, in general, the whole set of sites
of the lattice, but only those sites with two of the possible values of $s_i$.
In practical terms we can classify the cluster moves in two types:
Moves that change the number of occupied sites: Cluster N-sampling,
and moves in which the orientation of some of the sites can change: Cluster Orientation-sampling.
A full derivation of the procedures might be cumbersome, so we will just
include the steps and considerations required to understand the recipe of the algorithms.

\subsection{Cluster N-Sampling}

In these moves we consider only empty sites and occupied sites with one
(chosen at random) of the
six possible orientations,  $s_r$. These sites are named active sites.
We classify as passive (or blocked) those sites $k$ with $s_k \ne  s_r$ and  $s_k \ne  0$.
Passive sites are not modified
in these moves, and play the role of an external field. 
Taking into account the values of the interaction parameters, and in particular
that $\epsilon_{BB}=0$, one occupied active site interacts with another
occupied active site if (and only if) both sites are NN in the $s_r$ direction.
Using this definition of active sites the system can be equivalently seen
as a collection of one dimensional rows of sites (with PBC), i.e. 1D lattice
gas models under the influence of 
external fields.
The  terms of the Grand Canonical Hamiltonian that deppend on the active sites
 can be written as:
\begin{equation}
{\cal H}_{A} = - \epsilon \sum_{<ij>_{s_r}}^A \rho_i \rho_j - \sum_i^{A} \rho_i
\left[ \mu + \epsilon r N_1(i) + \epsilon r N_2(i) \right]
\end{equation}
where the first sum on the right hand side is carried out exclusively over pairs of active
sites which are NN along the  direction $s_r$. The second sum includes only
active sites. The variables $\rho_i$ take the values $0$  for empty sites and $1$ for 
occupied sites. $N_{1}(i)$ is the number of $A$ patches of the site $i$ that points to
a NN passive site, and $N_2(i)$ is the number 
of $A$ patches belonging to a NN passive site of $i$ that point to site $i$. Through  the
change of variables $\rho_i = 
( 1 + \sigma_i) / 2 $; ${\cal H}_A$, may be written as an Ising-like Hamiltonian:
\begin{equation}
{\cal H}_A -  {\cal H}_0 =  - \frac{\epsilon}{4} \sum_{<ij>_{s_r}}^A \sigma_i \sigma_j 
- \frac{\mu + \epsilon}{2} \sum_i^A\sigma_i 
- \sum_{i}^A \sigma_i \left[ \frac{\epsilon r}{2} N_2(i) + \left( \frac{\epsilon r}{2} -
\frac{\epsilon}{4} \right) N_1 (i) \right];
\label{mapI1}
\end{equation}
where ${\cal H}_0$ includes the terms that do not depend on the state of the active
sites. The new variables $\sigma_i$ can take the values $\pm 1$.  On the right hand
site of Eq. (\ref{mapI1}) the first term is the Ising-like interaction, the second 
plays the role of a global external field, and the last one includes the local external fields
that depend on the configuration of the passive sites.

Within this representation of the interactions of the active sites,  it is straighforward to
build up a cluster algorithm  
following  the Swendsen-Wang
procedure\cite{SwendsenWang} and its extensions in the presence  of external fields \cite{Landau_Binder}. The recipe of such
an algorithm goes as follows: (1)
Generate bonds between pairs, $\{i,j\}$, of active sites which are NN in the $s_r$ direction,
and that fulfill $s_i=s_j$ with probability:
\begin{equation}
B_{ij} = 1 - \exp \left[ - \epsilon/(2k_B T) \right];
\end{equation}
(2)  Consider separately each one of the clusters of active sites defined by the
previous bonds. Taking into account the effect of the external fields given in Eq.(\ref{mapI1}),  the new configuration is generated by assigning to all the lattice sites
in the cluster either $s=0$ (i.e. $\sigma=-1$);   or $s=s_r$ (i.e. $\sigma=1$) with probabilities 
$A(c,s)$ (where $c$ is the index for the $c$-th cluster)
fulfilling:

\begin{equation}
\frac{A(c,s_r)}{A(c,0)} = 
\exp \left\{   \frac{\mu + \epsilon }{k_B T} n_i(c) + 
\frac
{(r-1/2) \epsilon {\cal N}_1(c)  +\epsilon r {\cal N}_2(c)}
{  k_B T} \right\};
\end{equation}
where $n_i(c)$ is the number of lattice sites in the cluster $c$,  
${\cal N}_1(c)$ is the number of $A$ patches in the cluster $c$ that point to a NN blocked site, 
and ${\cal N}_2(c)$ is the number of A patches lying at blocked sites that point 
to a NN site in the cluster $c$.

\subsection{Cluster orientation sampling}

In these cluster moves two of the possible orientations, $s_a$, $s_b$, of the particles are chosen
as active directions, whereas the remaining four directions and the empty sites are classified as passive (or blocked) directions.
In the moves only active sites can modify their states (from $s_a$ to $s_b$ and vice versa). Therefore the number
of occupied sites will remain constant. Notice that the interaction between two active sites
that are NN through a passive direction is equal to $\epsilon_{BB}$ independently of their
respective orientations. In addition, the interaction between two NN sites, one being active and
the other passive can be modified in these moves only if they are NN through an active direction.

From these  features of the interaction potential, it follows that the only relevant
interactions in the proposed restricted sampling are those that take place between NN sites
through the active directions.  
As a consequence the system can be treated as a set of independent
layers with the topology of the square lattice (defined by the two active 
orientations), that can
contain active sites and two types of passive sites: empty and blocked sites.
The relevant terms of the potential energy on each layer take the form:
\begin{equation}
{\cal H}_A = - \epsilon \sum_{<ij>} \delta _{s_i,s_j} \delta_{s_i,\alpha_{ij}}  
- 
\epsilon r \sum_{[ik]}
  \delta_{s_i,\alpha_{ik}}  \left( 1 - \delta_{s_k,0} \right);
\end{equation}
where the $\delta$'s represent Kronecker delta functions, $<ij>$ indicates pairs
of NN active sites on the square lattice;  $[ik]$ stands for pairs of NN with
$i$ and $k$ being respectively an active and a passive site; and $\alpha_{lm}$ is the index of the direction
${\vec  r}_{lm}$.

Now, we describe the strategy to generate the cluster algorithm. In  previous papers
\cite{almarza2010,AlmarzaJCP2011}
we showed how the lattice patchy models defined on the square lattice at 
full occupancy
can be  mapped onto the two-dimensional lattice gas model. 
It can be shown
that it is also possible to carry out a mapping when some of the sites are blocked, the main
difference being that the effect of the blocked sites enters as a local external field.
\cite{Confined_SARR} This
mapping can be obtained through the plaquette procedures used in previous papers.
Taking into account plaquettes of four sites\cite{Noya,almarza2010}, and
defining on each plaquette $h_A$ as the sum of the potential energy 
contributions involving at least one active particle, it can
be shown that $h_A$
can be written in terms of a Potts-like interaction as:
\begin{equation}
h_A = h_0 - K  \sum_{<ij>}^{p}  \delta_{s_i,s_j} + 
K_{1} \sum_{[ik]}^{p}  \delta_{s_i,\alpha_{ik}} \left( 1 - \delta_{s_k,0} \right)
+K_{10} \sum_{[ik]}^{p}  \delta_{s_i,\alpha_{ik}} \delta_{s_k,0} ; 
\label{MapPotts1}
\end{equation}
where the superscript $p$ over the sums indicates that only interactions between
sites belonging to the plaquette are considered, $h_0$ is just an additive
constant which deppends on the configuration of the passive sites in the 
plaquette, but not on the state of the active sites, and finally $K$, $K_{10}$
and $K_0$  
deppend only on the energy parameters of the patchy model: $\epsilon$ and $r$.
Since ${\cal H}_A$ can be written as one half of 
the sum of the plaquettes energies $h_A$, we find
\begin{equation}
{\cal H}_A  - H_0 = - K  \sum_{<ij>} \delta_{s_i,s_j} + 
K_{1} \sum_{[ik]}  \delta_{s_i,\alpha_{ik}} \left( 1 - \delta_{s_k,0} \right)
+K_{10} \sum_{[ik]}  \delta_{s_i,\alpha_{ik}} \delta_{s_k,0} 
\label{MapPotts2}
\end{equation}
where
$H_0$ is an additive constant which does not deppend on the configuration
of the active sites,
$K \equiv (1/2-r) \epsilon$, $K_1 = K$, and $K_{10} = \epsilon/2$. On the
right and side of   Eq. (\ref{MapPotts2}) : the first term is a Potts $q=2$ interaction;
the second term includes the effective interaction of active sites with their NN
occupied passive sites (on the square lattice), whereas the last term represents
the effective interactions of active sites with their NN passive empty sites (on the square lattice).
The last two terms can be seen, as before, as local external fields.

Once the interaction between active sites has been described in terms of the Potts model,
it is straighforward to use the same strategy described for the cluster N-sampling.
Following Ref. (\onlinecite{Landau_Binder}),
the algorithm recipe is: 
Pairs of active sites, $(i,j)$ being NN (on the chosen square lattice)  that fulfill $s_i=s_j$ are bonded
with probability: 
\begin{equation}
B_{ij} = 1 - \exp \left[ - K /( k_B T ) \right].
\end{equation}
These bonds define clusters of active sites;  
and the  orientation of each of the clusters in the new configuration  is 
chosen from the active directions:
$s_{\alpha} = s_a, s_b$; 
with probabilities:
\begin{equation}
A(c,s_{\alpha}) \propto \exp \left[ - N^{(0)}(s_\alpha) \frac{\epsilon}{2k_B T} 
- N^{(1)}(s_\alpha) \frac{(1/2-r) \epsilon}{ k_BT} \right];
\end{equation}
where $N^{(0)}(s_{\alpha})$ is the number of A patches belonging to the cluster that point
to empty blocked NN sites when the particles of the cluster are oriented in direction $s_{\alpha}$,
and $N^{(1)}(s_{\alpha})$ is the number of A patches belonging to the cluster pointing to
occupied blocked NN sites when the cluster is oriented in direction $s_{\alpha}$.

\end{document}